\documentclass[epj]{svjour}
\usepackage{psfig}
\begin{document}

\title{Polarization observables in high-energy deuteron photo\-disintegration
within the Quark-Gluon Strings Model\thanks{ Supported by DFG, RFFI and
Forschungszentrum J\"ulich.}}

\author{V.Yu.~Grishina $^a$, L.A.~Kondratyuk $^{b,c}$, W.~Cassing $^d$,
E.~ De~Sanctis $^e$, M. Mirazita $^e$, F. Ronchetti $^e$ and P.~Rossi $^e$ \\}
\institute{$^a$ Institute for Nuclear Research, 60th October
Anniversary Prospect 7A, 117312 Moscow, Russia\\
$^b$ Institute for Theoretical and Experimental Physics, B.\
  Cheremushkinskaya 25, 117259 Moscow, Russia \\
$^c$ IKP, Forschungszentrum J\"ulich, D-52425 J\"ulich,Germany\\
$^d$ Institute for Theoretical Physics,  University of
Giessen, Heinrich-Buff-Ring 16, D-35392 Giessen, Germany\\
$^e$ Frascati National Laboratories, INFN, CP 13, via E.
Fermi, 40; I-00044, Frascati, Italy}
\date{Received: date / Revised version: date}

\abstract{Deuteron two-body photodisintegration is analysed within
the framework of the Quark-Gluon Strings Model. The model describes
fairly well the recent experimental data from TJNAF in the
few GeV region. Angular distributions at
different $\gamma$-energies are presented and the effect of a
forward-backward asymmetry is discussed. New results from the QGSM
for polarization observables from 1.5 -- 6 GeV are presented and
compared with the available data.}


\authorrunning{V. Yu. Grishina et al.}

\titlerunning{Polarization observables...}
\maketitle

In a recent paper \cite{Ref-1} we have investigated high-energy
deuteron photodisintegration within the framework of the
Quark-Gluon Strings Model (QGSM). The QGSM - proposed by Kaidalov
\cite{Kaidalov} - is based on two ingredients: i) a topological
expansion in QCD and ii) a space-time picture of the interactions
between hadrons that takes into account the confinement of quarks.
In a more general sense the QGSM can be considered as a
microscopic (nonperturbative) model of Regge phenomenology for the
analy\-sis of exclusive and inclusive hadron-hadron and
photo-hadron reactions on the quark level. Within the QGSM the
deuteron photodisintegration amplitude $T(\gamma d \to pn)$ can be
described in first approximation by the planar graph with three
valence quark exchange in $t$ (or $u$)-channels, which corresponds
to a nucleon Regge trajectory. The intermediate $s$ channel will
consist of a $6q$ string (or color tube) with $1q$ and $5q$ states
at the ends. Assuming that all the intermediate quark clusters
have minimal spins and the $s$ channel helicities in the
quark-hadron and hadron-quark transition amplitudes are conserved,
we can reconstruct the spin structure of the amplitude $T(\gamma d
\to pn)$ as \cite{Ref-1}
\begin{eqnarray}
\lefteqn{ \langle p_3,\lambda_{p}; p_4,\lambda_{n} |
\hat{T}\left(s, t\right)|
p_2,\lambda_{d};p_1,\lambda_{\gamma}\rangle =}\nonumber \\ &&\bar
u_{\lambda_p}(p_3) \hat {\epsilon}_{\lambda_{\gamma}}
\left({A(s,t) \hat{p}_3 +B(s,t) m}\right) \hat
{\epsilon}_{\lambda_{d}} v_{\lambda_n}(p_4)\ . \label{spin1}
\end{eqnarray}
The ratio $R=B(s,t)/A(s,t)$ is fixed by the model to the interval
$R=1\div 2$; calculations will be presented for the limiting
values $R=1$ and $R=2$ (see below).

In Ref. \cite{Ref-1} we have analyzed deuteron photodisintegration
within the framework of the QGSM employing nonlinear Regge
trajectories. Parameters have been fixed by a previous analysis of
$pp$ data (i.e.  the reaction $pp \to d \pi^+$) and the TJNAF data
at $\Theta_{\mathrm{c.m.}} = 36^{\circ}$. We have found that the
QGSM provides a reasonable description of all new TJNAF data
\cite{Shulte}--\cite{Ref-3} on deuteron photodisintegration at
large momentum transfer $t$ and that the energy dependence of
${\mathrm{d}} {\sigma}/{\mathrm d}t$ at $\Theta_{\mathrm{c.m.}} =
36\div 90^{\circ}$ gives new evidence for a nonlinearity of the
Regge trajectory $\alpha_N(t)$. The best agreement with the data
can be achieved using the -- QCD motivated -- logarithmic form of
the Regge trajectory \cite{Ref-1},
\begin{equation} \label{nonlin} \alpha_N(t)
= \alpha_N(0)- (\gamma \nu) \ln (1 - {t}/{T_B}). \end{equation}
Evidently the QGSM predicts that $d\sigma/dt $ at fixed c.m.
angles will decrease faster than any finite power of $s$ and at
sufficiently large energies the perturbative regime will become
dominant. Therefore, it is very important to have new data  at
larger energies to further check the energy behaviour of
$d\sigma/dt $ as obtained from the QGSM.
\begin{figure}[h]
\begin{center}
\centerline{\psfig{file=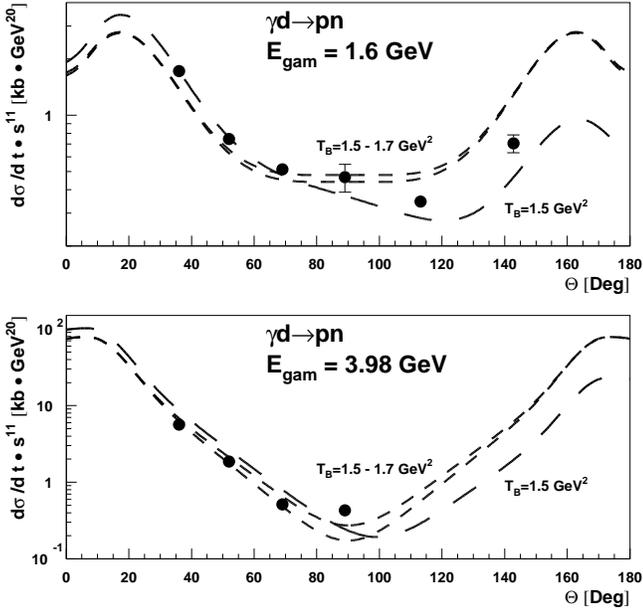,width=8.5cm}}
\caption{Differential cross section for the reaction $\gamma d\to
pn$ (multiplied by $s^{11}$) as a function of the c.~m. angle for
$E_{\gamma}=1.6$~GeV and 3.98~GeV \protect\cite{Ref-1}.  The
experimental data are from Ref. \cite{Ref-3}. The dashed curves
are calculated within the QGSM with logarithmic Regge trajectories
(2) using $T_B = 1.5$ and 1.7 GeV$^2$ (cf. \cite{Ref-1}). The
long-dashed curve presents the result of calculations which take
into account the interference of the isoscalar and isovector parts
of the $\gamma d\to pn$ amplitude (see text).} \label{fig:dsgb1}
\end{center}
\end{figure}
\begin{figure}[h]
\begin{center}
\centerline{\psfig{file=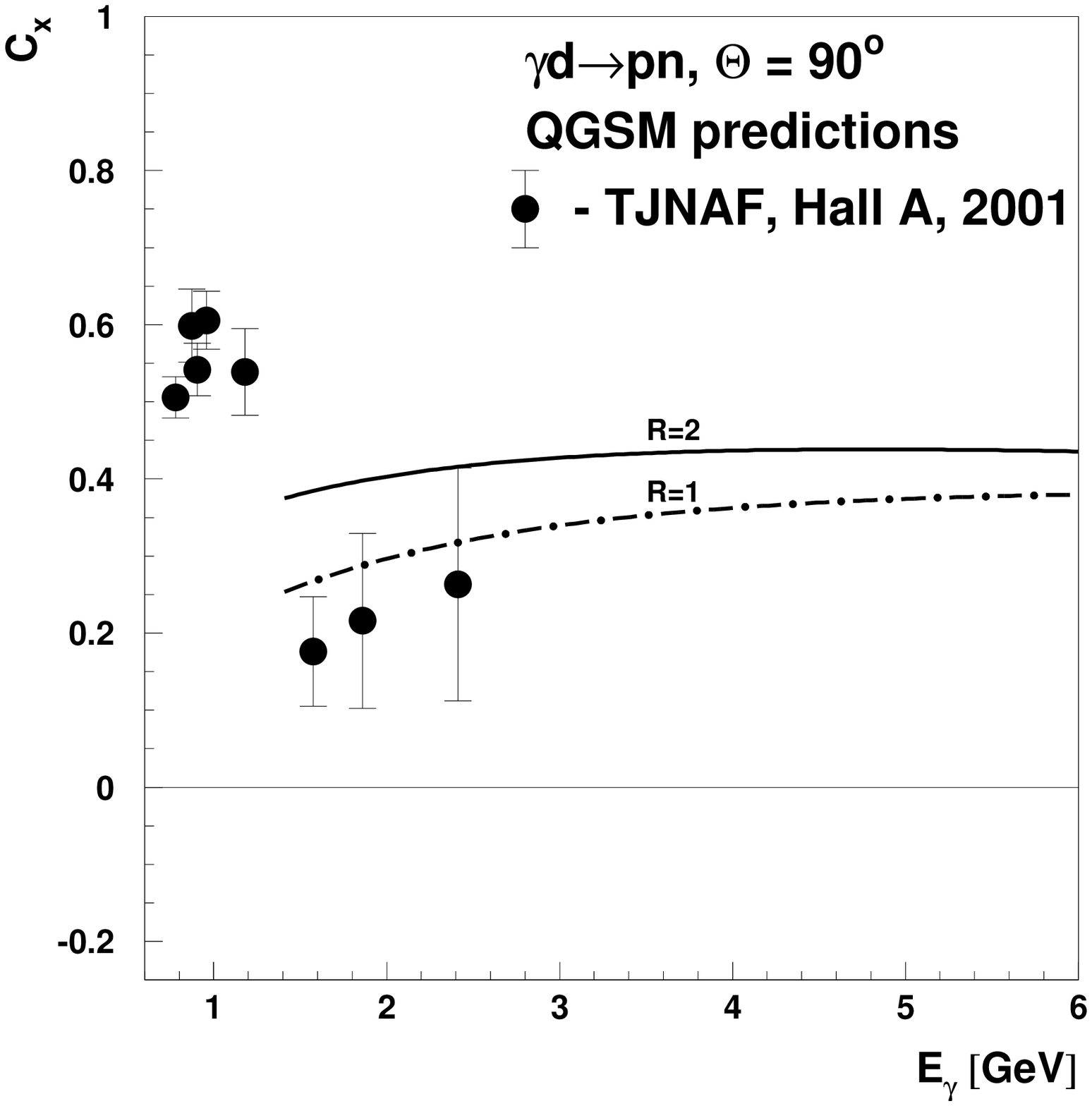,width=8.5cm,height=7.5cm}}
 \caption{Polarization transfer $C_x$ for circularly polarized photons. The
solid and dash-dotted curves correspond to $R=2$ and 1,
respectively. The experimental data are from \cite{Ref-4} and have
been corrected for spin rotation due to the lab. -- c.m.
transformation.} \label{fig:gdpncx}
\end{center}
\end{figure}
\begin{figure}[h]
\begin{center}
\centerline{\psfig{file=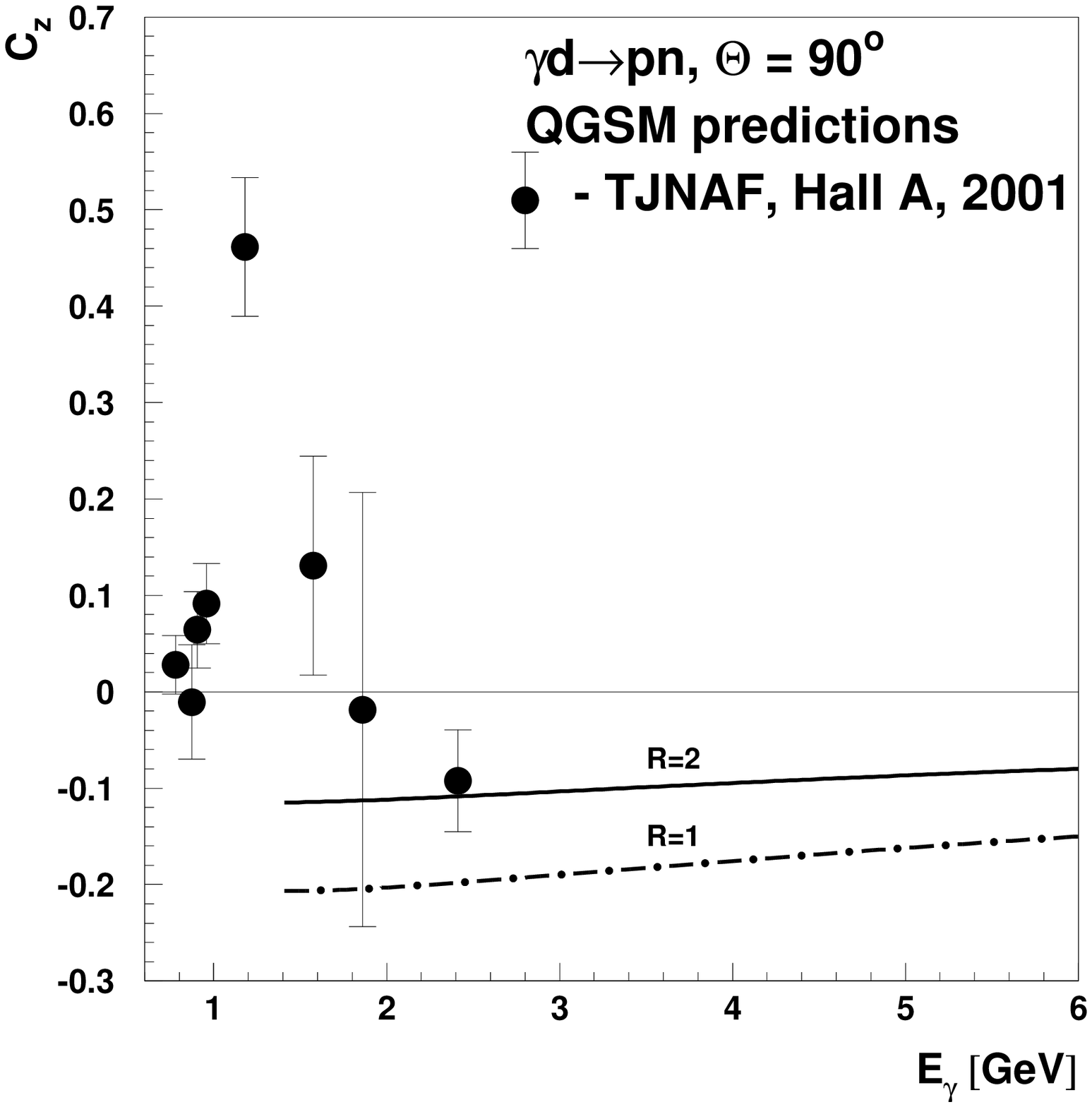,width=8.5cm,height=8cm}}
\caption{ Polarization transfer $C_z$ for circularly polarized
photons. The solid and dash-dotted curves correspond to $R=2$ and
1, respectively. The experimental data are taken from \cite{Ref-4}
and have been corrected for spin rotation due to the lab. -- c.m.
transformation.} \label{fig:gdpncz}
\end{center}
\end{figure}
\begin{figure}[h]
\begin{center}
\centerline{\psfig{file=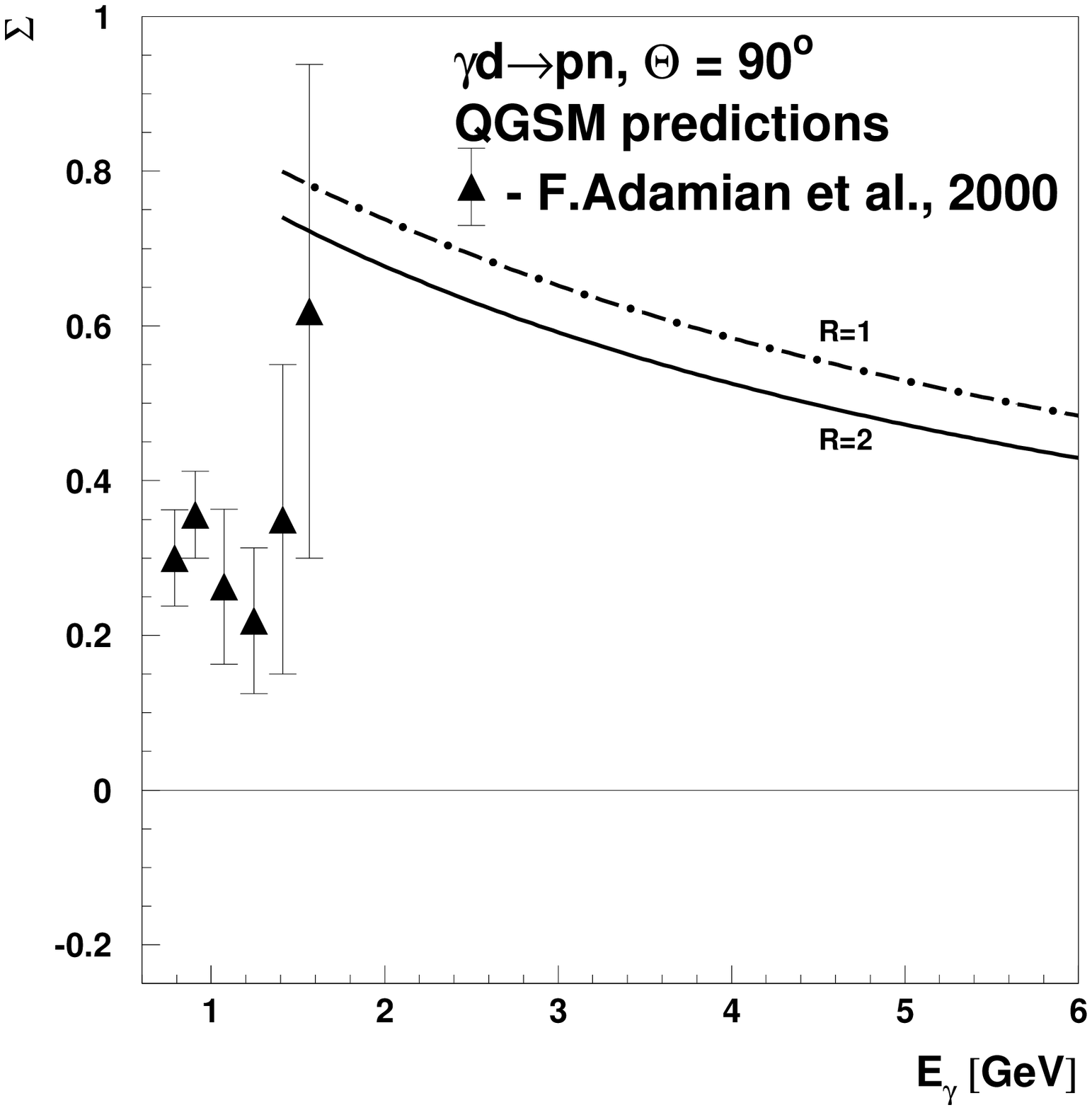,width=8.5cm,height=8cm}}
\caption{The asymmetry $\Sigma (90^{\circ})$  for linearly
polarized photons as a function of the photon energy. The solid
and dash-dotted curves correspond to $R=2$ and 1, respectively.}
\label{fig:gdpnsigma90}
\end{center}
\end{figure}

We have also investigated the angular dependence of the cross
section at different energies. In Fig. \ref{fig:dsgb1} we present
the angular  dependence of ${d\sigma}/{dt}\cdot s^{11}$ at two
energies ($E_{\gamma}$ =1.6 and 3.98 GeV) for the logarithmic
Regge trajectory (2). The two dashed curves have been calculated
assuming isovector photon dominance \cite{Ref-1}. In this case we
get a forward-backward symmetry of the differential cross section.
At 1.6 GeV the calculated angular distribution has a dip for
$\Theta_{\mathrm{c.m.}}$ =0$^{\circ}$ and 180$^{\circ}$ which is
related to the choice of the ratio $R=2$. This dip is absent for
$R=1$. The long-dashed curves in Fig. 1 take into account the
forward-backward asymmetry as discussed below.

We recall, that a forward-backward asymmetry arises from the
interference of two amplitudes which describe the contribution of
isovector ($\rho$ like) and isoscalar ($\omega $ like) photons. In
this case the differential cross section can be written as
\begin{eqnarray}
&&\displaystyle \frac{d\sigma^{\rho+\omega}_{\gamma d\to pn}}{d t}
= \nonumber \\ &&\frac{1}{64\,\pi s}\
\frac{1}{(p_{\gamma}^{\mathrm{cm}})^2}\ \left| \langle \lambda_p;
\lambda_n \left|\hat{T}{}^{\rho}(s,t)+\hat{T}^
{\omega}(s,t)\right|\lambda_d ;\lambda_{\gamma} \rangle \right.-
\nonumber\\ && \vphantom{ \frac{1}{64\,\pi s}
\frac{1}{(p_{\gamma}^{\mathrm{cm}})^2}} \left. \langle \lambda_p;
\lambda_n \left|\hat{T}{}^{\rho}(s,t)-\hat{T}^
{\omega}(s,t)\right|\lambda_d ;\lambda_{\gamma} \rangle \right|^2
. \label{eq:sigtas}
\end{eqnarray}
Using the vector dominance model (VDM) we adopt
\begin{equation}
\hat{T}^{\omega}(s,t)=\hat{T}^{\rho}(s,t)/\sqrt{8}, \hspace{0.4cm}
\hat{T}^{\omega}(s,u)=\hat{T}^{\rho}(s,u)/\sqrt{8}. \end{equation}
The data in Fig. 1 at 1.6 GeV provide evidence for a
forward-backward asymmetry because the differential cross section
at backward angles is  smaller than  for the corresponding angles
in the forward region. The predictions of the QGSM model with
$\rho -\omega$  interference are in qualitative agreement with the
published data \cite{Ref-3}(long-dashed lines). They are also in
good agreement with the new preliminary data from the CLASS
collaboration at TJNAF \cite{Ref-2,Ronchetti}.

Complementary information to the differential cross sections is
provided by polarization observables that should give important
tests of nonperturbative calculations in the intermediate energy
regime. Data for recoil polarizations have been published recently
\cite{Ref-4}. Existing Meson-Baryon Models (MBM) fail to describe
the data for the induced polarizations, which are surprisingly
small for energies above about 1 GeV. Moreover, the polarization
transfer data are inconsistent with hadron helicity conservation
(HHC), which is generally expected for PQCD.

We present here the polarization observables as calculated
within the QGSM. For the definitions of these observables in terms
of helicity amplitudes we refer the reader to Ref.
\cite{Barannik}. We find that\\ i) the induced polarization $P_y$
vanishes at $\Theta_{c.m.} =90^{\circ}$, but is different from 0
for $\Theta_{c.m.} \neq 90^{\circ}$;\\ ii) the polarization
transfers $C_x$ and $C_z$ for circularly polarized photons do not
vanish at $\Theta_{c.m.} =90^{\circ}$: $C_x \simeq 0.25\div 0.35$
(Fig. \ref{fig:gdpncx}) and $C_z \simeq -0.1\div -0.2$ at 1.5-2.5
GeV (Fig. \ref{fig:gdpncz}) ;
\\ iii) the polarized photon asymmetry $\Sigma$ is about 0.7 at
$\Theta_{c.m.} =90^{\circ}$ and $E_{\gamma}$=1.5 GeV (Fig.
\ref{fig:gdpnsigma90} ) and drops smoothly with $E_\gamma$.

We observe that for $E_{\gamma} \geq 1.5$ GeV the calculated
values of $C_x$ and $C_z$ are in agreement with the new TJNAF
data. We note that the term $\sim B(s,t)$ in (1) violates
chirality; consequently our results for $C_x$ and $C_z$ are
different from HHC prediction
($C_x=C_z=0$), which is  an essential property of PQCD. According
to Brodsky and Hiller \cite{Brodsky}  we should have $\lambda_d =
\lambda_p + \lambda_n$ within PQCD independently of
$\lambda_{\gamma}$. Assuming that -- in the scaling limit -- the
transverse deuteron helicities are suppressed as compared to the
longitudinal ones the asymmetry for linearly polarized photons
\begin{equation}
\Sigma(\theta) = (d\sigma_{||} - d\sigma_{T})/
(d\sigma_{||} + d\sigma_{T})
\end{equation} should at $90^{\circ}$ approach the value \cite{Nagornyi}:
\begin{equation} \Sigma \simeq -2Re(T_{+-}^{10}T_{-+}^{10*})/ (|T_{+-}^{10}|^2
+ |(T_{-+}^{10}|^2). \end{equation} Nagornyi et al.
\cite{Nagornyi} predicted -- using the axial symmetry
$T_{+-}^{10}(90^{\circ})=T_{-+}^{10}(90^{\circ})$ --  that $\Sigma
(90^{\circ}$) should approach --1. We note, however, that the
condition $T_{+-}^{10}(90^{\circ})= T_{-+}^{10}(90^{\circ})$ is
valid only for 'isoscalar' photons, where the isospin function is
antisymmetric. But in case of 'isovector' photons the isospin
function is symmetric and due to the Pauli principle we have
$T_{+-}^{10}(90^{\circ})=-T_{-+}^{10}(90^{\circ})$. Furthermore,
according to the VDM the isovector photon couples to hadrons more
strongly that the isoscalar photon. Thus one expects that in case
of HHC the polarization $\Sigma
(90^{\circ})$  should not be very different from +1.

As seen from Fig. 4 the QGSM predicts a slow decrease of $\Sigma
(90^{\circ})$ with photon energy from 0.7--0.8 at 1.5 GeV to
0.4--0.5 at 5--6 GeV. For photon energies below 1.5 GeV the
polarizations $C_x$ and $C_z$ show quasi-resonance structures at
$\sqrt{s}=2.7$ GeV and $\sqrt{s}=2.9$ GeV, respectively. Such
resonance structures are conceptually out of the range of the
QGSM.  It is interesting to note that there are similar structures
at exactly the same $\sqrt{s}$ in $\Delta \sigma _{L}$ for elastic
$pp$ scattering \cite{Auer}. We speculate that these
quasi-resonance structures might be related to the thresholds
$\rho NN$ and $\rho N \Delta$.

In summarizing this contribution we have presented angular
distributions at different $\gamma$-energies (1.68 GeV and 3.98
GeV) for deuteron photodisintegration within the QGSM and
discussed the effect of a forward-backward asymmetry. In addition
to Ref. \cite{Ref-1} new results from the QGSM for polarization
observables from 1.5 -- 6 GeV (cf. Figs. 2 -- 4) have been
calculated and compared to the available data.

\vspace{0.5cm} We are grateful to Ronald Gilman for sending us the
experimental data on the polarizations $C_x$ and $C_z$ corrected
for spin rotation due to the lab. -- c.m. transformation.

\end{document}